\documentclass{iau}
\usepackage{graphicx}
\usepackage{hyperref}

\title[VLTI/MIDI survey of young Sun-like stars] 
{A mid-infrared interferometric survey of the planet-forming region around young Sun-like stars}

\author[J\'{o}zsef Varga \& al.]   
{J\'{o}zsef Varga$^1$,
   		  P\'eter \'Abrah\'am$^1$,
          Lei Chen$^1$,
          Thorsten Ratzka$^2$, 
          K. \'E. Gab\'anyi$^{1,3}$, and  
          \'A. K\'osp\'al$^{1,4}$ 
          }

\affiliation{$^1$Konkoly Observatory, Research Centre for Astronomy and Earth Sciences, Hungarian Academy of Sciences, Konkoly Thege Mikl\'os \'ut 15-17., H-1121 Budapest, Hungary \\ email: {\tt varga.jozsef@csfk.mta.hu} \\[\affilskip]

$^2$Institute for Physics/IGAM, University of Graz, Universit\"atsplatz 5/II, 8010, Graz, Austria 

$^3$MTA-ELTE Extragalactic Astrophysics Research Group, P\'azm\'any P\'eter s\'et\'any 1/A H-1117 Budapest Hungary

$^4$Max Planck Institute for Astronomy, K\"onigstuhl 17, 69117 Heidelberg, Germany

}

\pubyear{2019}
\volume{345}  
\jname{Origins: from the Protosun to the First Steps of Life}
\editors{Bruce G. Elmegreen, L. Viktor T\'oth, Manuel G\"udel, eds.}
\begin{document}

\maketitle

\begin{abstract}
We present our results from a mid-infrared interferometric survey targeted at the planet-forming region in the circumstellar disks around low- and intermediate-mass young stars. Our sample consist of 82 objects, including T Tauri stars, Herbig Ae stars, and young eruptive stars. Our main results are: 1) Disks around T Tauri stars are similar to those around Herbig Ae stars, but are relatively more extended once we account for stellar luminosity. 2) From the distribution of the sizes of the mid-infrared emitting region we find that inner dusty disk holes may be present in roughly half of the sample. 3) Our analysis of the silicate spectral feature reveals that the dust in the inner $\sim$1~au region of disks is generally more processed than that in the outer regions. 4) The dust in the disks of T Tauri stars typically show weaker silicate emission in the N band spectrum, compared to Herbig Ae stars, which may indicate a general difference in the disk structure. Our data products are available at VizieR, and at the following web page: \url{http://konkoly.hu/MIDI_atlas}.
\end{abstract}

\firstsection 
\section{Introduction}

Circumstellar disks around young stellar objects (YSOs) are the places where planetary systems are born. These disks have a lifetime of about 10 million years. During this relatively short time disk material is accreted by the central star. Dust grains grow and stick together, gradually building planets. The protoplanetary disk eventually disperses, leaving a debris disk from the remaining material. Important aspects of disk and planetary evolution, e.g., dust evolution or the clearing of the dusty disk, are still not well understood, since they can only be tackled with high angular resolution observations. 

Long-baseline interferometers at mid-infrared (mid-IR) wavelengths can reach angular resolution around $\sim$10~mas. 
With mid-IR interferometry we can probe the structure of the warm dust, located at typically $1-10$~au from the star (in the planet forming region). Additionally, spectrally resolved mid-IR interferometric observations provide an insight to the composition and spatial distribution of warm dust throughout the disk. 

Here we present an overview of our mid-IR interferometric survey of young Sun-like stars, and highlight some results from the statistical analysis of our sample.

\section{Overview}

The now decommissioned Mid-infrared Interferometric Instrument (MIDI, \cite[Leinert et al. 2003]{midi}) at the Very Large Telescope Interferometer (VLTI) left a rich heritage of high spatial resolution observations of planet-forming disks. It provided spectrally resolved interferometric observations in the N band ($7.5-13\ \mu$m), covering the silicate spectral feature.  However, large part of the data remained unpublished, partly because the reduction and the interpretation of IR interferometric data is not always straightforward. In \cite{Varga2018} we published significant part of these data, focusing on low-mass YSOs (T Tauri stars), the younger siblings of the Sun. We collected MIDI observations on 45 T Tauri stars, but for comparison we also included in our sample 26 Herbig Ae stars, and 11 young eruptive stars. In total we selected 82 objects with $\sim$600 interferometric measurements over $\sim$200 nights between 2003 and 2015. The data were reduced homogeneously using the standard MIDI pipeline, EWS (e.g., \cite[Chesneau 2007]{Chesneau2007}), with the help of the routines of \cite{Menu2015}. The resulting total spectra (representing the whole disk's emission), correlated spectra (the emission from the inner disk region, where the size of this region depends on the baseline length), spectrally resolved visibilities, and differential phases are published online in the form of an atlas at \url{http://konkoly.hu/MIDI_atlas}, and also at VizieR (\url{http://cdsarc.u-strasbg.fr/viz-bin/cat/J/A+A/617/A83}).

To estimate the extent of the warm dusty part of the disks, for each object we fitted a simple geometric disk model to the $10.7\,\mu$m correlated fluxes. The result of the modelling is the half-light radius of the mid-IR emitting region, defined as the radius at which the integrated intensity is half of the total flux density.

\begin{figure}[b]
\begin{center}
 \includegraphics[width=12.5cm]{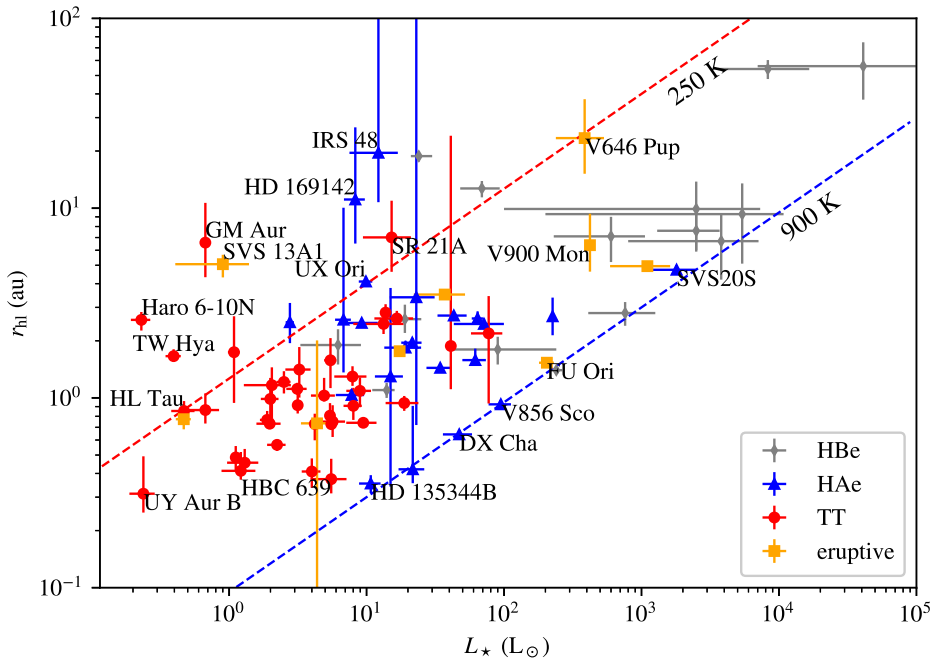}
 \caption{Size (half-light radius) of the mid-IR emitting region of circumstellar disks as a function of the stellar luminosity. Points for T Tauri (red circle), Herbig Ae (blue triangle), and young eruptive (orange square) stars are from our work. Data for Herbig Be stars (grey diamonds) are from \cite{Menu2015}. The radii that correspond to temperatures of $250$~K and $900$~K for optically thin gray dust are plotted with red and blue dashed lines, respectively.}
   \label{fig:size}
\end{center}
\end{figure}

\section{Implications}

{\underline{\it Size of the mid-IR emitting region}}.
Disks around T Tauri and Herbig Ae/Be stars are illuminated by the central star, and re-emit this radiation at IR wavelengths. Thus, there is a correlation between the IR appearance of the disks and the luminosity of the central star. In the near-IR this correlation is quite strict (\cite[Eisner et al. 2007]{Eisner2007}), indicating that the near-IR emitting region is usually confined to a region near the dust sublimation radius, which is almost entirely determined by the star (as the place where the temperature is $\sim$1500~K). In Fig.\,\ref{fig:size} we plot the size-luminosity relation from our data. The correlation between the mid-IR emitting region and the stellar luminosity is much weaker compared to the near-IR. This reflects the structural diversity of the mid-IR emitting region. The radii that correspond to temperatures of 250~K and 900~K for optically thin gray dust are also plotted. Thus, the location of points in the figure can be connected to the typical temperature of the dust. Points located above the 250~K line indicate disks that are relatively cold and oversized. This can be caused by the inner clearing of the disk, and it is usually observed among transitional disks (e.g., TW Hya, GM Aur, Haro 6-10N, HD 169142). Several Herbig Ae stars are located close to the 900~K line, indicating relatively compact disks. In contrast, T Tauri stars preferentially lie closer to the 250~K line, suggesting that disks around T Tauri stars are relatively more extended and colder than those around Herbig Ae stars, once we account for stellar luminosity. Inner disk clearing can be caused by e.g., photoevaporation, grain growth, and planet formation. We simulated a population of disk models with radiative transfer modeling. The model population had two groups: in the first group the dusty disk begins at the sublimation radius, while in the other it starts further out due to an inner hole. By comparing the size distribution of our objects with the disk models, we found that roughly half of our objects can have an inner dusty disk hole.

\begin{figure}[t]
\begin{center}
     \includegraphics[width=10cm,trim={0 0.55cm 0 0.37cm},clip]{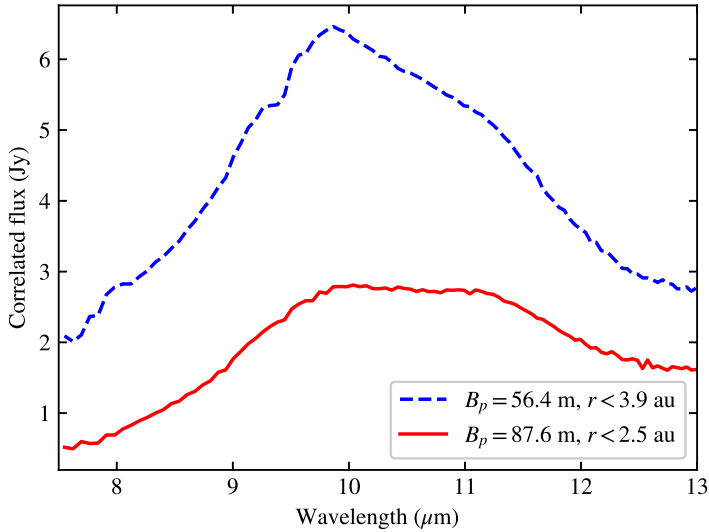}
 \caption{VLTI/MIDI correlated spectra of GW Ori showing the N band silicate spectral feature. The blue dashed line represents the emission from the inner $3.9$~au, while the red solid line represents the emission from the inner $2.5$~au.}
   \label{fig:spec}
\end{center}
\end{figure}

{\underline{\it Dust mineralogy}}. The first important step towards the formation of planetesimals is dust processing. It is known from mid-IR spectroscopy of the N band silicate feature that the original small amorphous dust grains interact with each other, grow and/or partly transform into crystalline species (\cite[van Boekel et al. 2003, Przygodda et al. 2003]{vanBoekel2003,Przygodda2003}). \cite{vanBoekel2004} studied MIDI spectra of Herbig Ae/Be stars, and revealed that primordial dust is mainly found in the outer disk, while more evolved dust is located closer to the star. Here we study dust processing both in T Tauri and Herbig Ae stars from our sample. The shape of the N band silicate feature reflects the composition and size of the dust grains. For small ($\sim$$0.1\,\mu$m) amorphous grains the feature has a triangular shape peaking at $9.8\,\mu$m, while for larger ($\sim$$1-2\,\mu$m) grains the spectrum will show a plateau between 10 and $11\,\mu$m. Crystalline silicates have smaller sharp features at several wavelengths (e.g., at $9.2$, $10.5$ and $11.3\,\mu$m). In Fig.\,\ref{fig:spec} we show two correlated spectra of the T Tauri star GW Ori as an example. The spectra represent the emission from the disk inside $r=3.9$~au ($9.8$~mas, dashed blue line), and inside $r=2.5$~au ($6.3$~mas, solid red line), respectively. The $r<2.5$~au spectrum show a plateau characteristic of large grains, while in the $r<3.9$~au spectrum a triangular peak centered at $9.8\,\mu$m appears. It clearly indicates dust processing, with larger grains located closer to the star, in agreement with the findings of \cite{vanBoekel2004}. We found a similar trend for most objects in our sample. The different radial distribution of small and large grains indicate that grain processing mostly happens in the inner $\sim$1~au region of planet-forming disks, where the temperature and density are higher, and the dynamical timescales are smaller. We also found that the objects showing the strongest silicate emission in their spectra are all Herbig Ae stars. T Tauri stars typically have weaker features, indicating a possible difference in the disk structure. As the silicate emission feature originates from the disk surface, disks with larger surface area in the mid-IR emitting region are expected to show a stronger silicate feature. The size of the surface heavily depends on the disk flaring. Thus, we propose that T Tauri disks may be less flared than Herbig Ae disks.

\acknowledgements This project has received funding from the European Research Council (ERC) under the European Union's Horizon 2020 Research and Innovation programme under grant agreement No. 716155 (SACCRED).

\begin{discussion}

\discuss{Johnstone} {Beautiful observations. How much degeneracy remains in the
interpretation? You provide very nice results but what else is needed to really
confirm that these observations really lead to the physical trends suggested?}

\discuss{Varga} {Due to sparse baseline coverage, many different disk models can fit the interferometric data. By using simple models we can get robust estimates, like the half-light radius. An other ambiguity lies in the interpretation of the  
silicate spectral feature, as both large amorphous grains and crystalline grains can produce a similar feature shape. To tackle these issues, better baseline coverage and higher spectral resolution is needed.}

\end{discussion}

\end{document}